\documentclass[aps, prb, twocolumn, amssymb, amsmath, showpacs, superscriptaddress]{revtex4-1}
\usepackage{bm}
\usepackage{times}
\usepackage{graphicx}
\usepackage{makecell}
\usepackage{color}
\usepackage{dcolumn}
\usepackage[colorlinks=true, letterpaper=true, pdfstartview=FitV, linkcolor=blue, citecolor=blue, urlcolor=blue]{hyperref}
\usepackage{appendix}
\usepackage{mathrsfs}

\usepackage{helvet}

\usepackage{pifont}

\usepackage{hyperref}

\begin{document}

\title{Hierarchy of Electromagnetic Gauge Field Quantum Geometries}

\author{Bin Wang}
\affiliation{College of Physics and Optoelectronic Engineering, Shenzhen University, Shenzhen 518060, China}
\author{Fuming Xu}
\affiliation{College of Physics and Optoelectronic Engineering, Shenzhen University, Shenzhen 518060, China}
\affiliation{Quantum Science Center of Guangdong-Hongkong-Macao Greater Bay Area, Shenzhen 518045, China}
\author{Lei Zhang}
\email{zhanglei@sxu.edu.cn}
\affiliation{State Key Laboratory of Quantum Optics Technologies and Devices,
Institute of Laser Spectroscopy, Shanxi University, Taiyuan 030006, China}
\affiliation{Collaborative Innovation Center of Extreme Optics, Shanxi University, Taiyuan 030006, China}
\author{Jian Wang}
\email{jianwang@hku.hk}
\affiliation{College of Physics and Optoelectronic Engineering, Shenzhen University, Shenzhen 518060, China}
\affiliation{Quantum Science Center of Guangdong-Hongkong-Macao Greater Bay Area, Shenzhen 518045, China}
\affiliation{Department of Physics, The University of Hong Kong, Pokfulam Road, Hong Kong, China}

\begin{abstract}
The conventional Bloch quantum geometric tensor (QGT) describes
electric-dipole geometry for spatially uniform perturbations and
vertical Bloch transitions. We develop a broader framework linking
quantum geometry to local and nonlocal electromagnetic response
in gauge-field space, placing this tensor within a hierarchy of
electric and magnetic multipole quantum geometries.
The intrinsic antisymmetric adiabatic response in every
electromagnetic channel is a Berry curvature in the space of
its conjugate fields. The field-dependent projector uniquely fixes the QGT for that
channel, pairing the response curvature with a quantum metric. This construction extends to spatially nonuniform perturbations,
whose finite momentum transfer induces nonvertical Bloch transitions. We explicitly construct the current--current QGT connecting
Bloch states at $\mathbf{k}\pm\mathbf{q}/2$, providing a unified
metric and curvature description of nonlocal current response. In the doubled Brillouin zone, tangential derivatives
generate conventional momentum-space QGT multipoles, whereas
normal derivatives connect finite-momentum geometry to uniform
electric- and magnetic-multipole-pair QGTs. We derive this connection through second order in $\mathbf{q}$. The finite-momentum curvature
governs the intrinsic nonlocal Hall response, while an
inverse-frequency moment of symmetric optical absorption probes
the index-symmetric metric sector.
\end{abstract}

\maketitle

\noindent \textit{\textcolor{blue}{Introduction.}}---
Quantum geometry has become a central organizing principle for electronic response. The quantum geometric tensor (QGT) combines the Berry curvature, which captures phase structure, and the quantum metric, which measures distance in projective Hilbert space \cite{ProvostVallee1980,Berry1984,XiaoChangNiu2010}. In crystalline solids, these quantities govern localization and Wannier functions \cite{RestaSorella1999,Resta2011,Kolodrubetz2017,Marzari2012}, optical spectral weight and dichroism \cite{SouzaVanderbilt2008,OzawaGoldman2018,Ahn2022}, superfluid transport \cite{PeottaTorma2015,Julku2016,Liang2017}, orbital magnetic response \cite{GaoYangNiu2014,GaoYangNiu2015,Piechon2016}, and linear and nonlinear Hall effects \cite{XiaoChangNiu2010,SodemannFu2015}. Their momentum derivatives further generate Berry-curvature and quantum-metric multipoles \cite{SodemannFu2015,GaoXiao2019}.

\begin{figure}[t!]
\includegraphics[width=1.0\columnwidth]{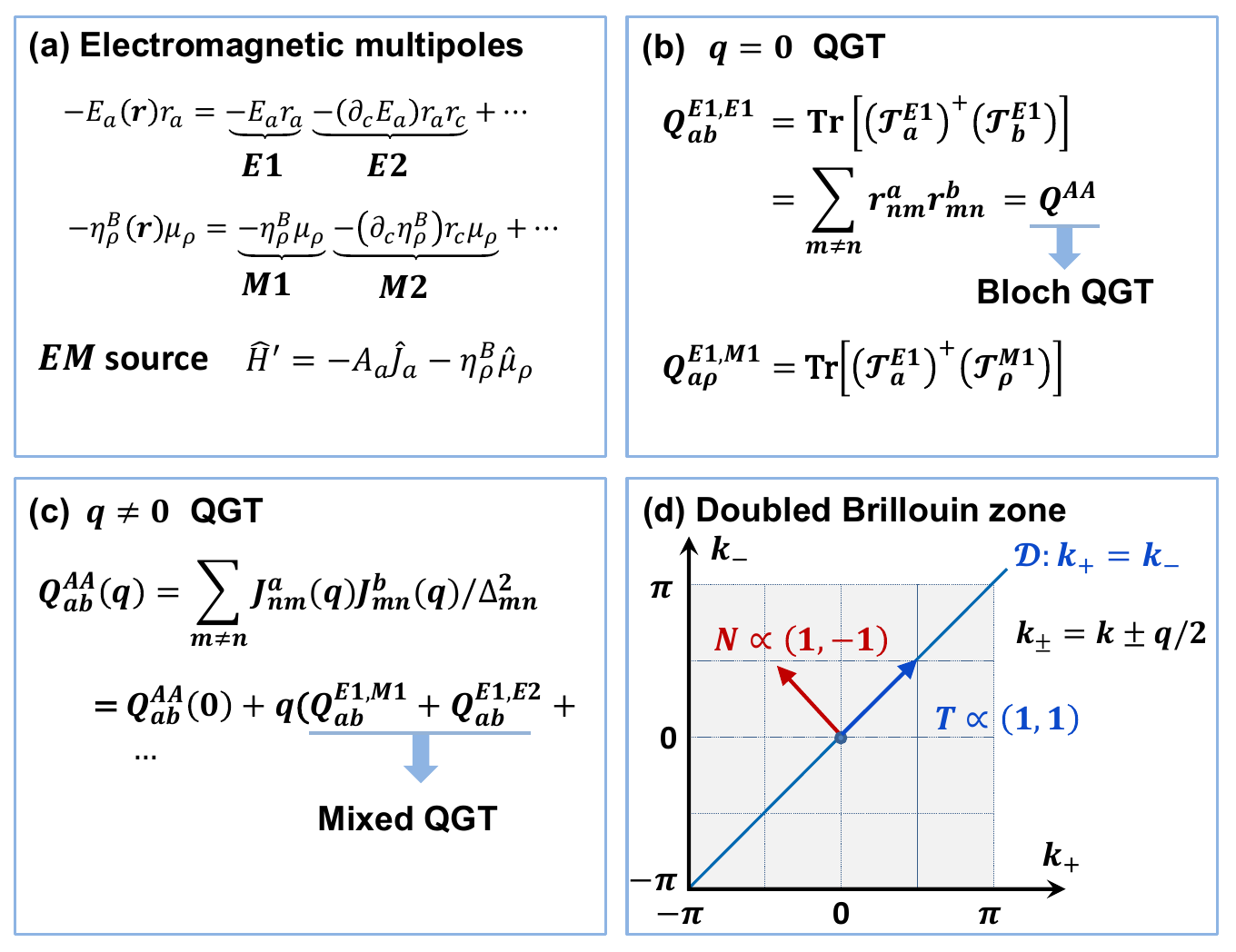}
\caption{Hierarchy of electromagnetic quantum geometries.
(a) Electric and magnetic multipole couplings.
(b) Uniform electric-dipole and mixed electric--magnetic QGTs.
(c) The finite-momentum current--current QGT connects uniform
multipole-pair geometries through its normal expansion.
(d) Tangential and normal variations in the doubled Brillouin
zone generate momentum-space multipoles and electromagnetic
multipole-pair geometries, respectively.}
\label{fig1}
\end{figure}

Electromagnetic interactions are conventionally organized by electric and
magnetic multipoles, denoted by $E1,E2,\ldots$ and $M1,M2,\ldots$ (Fig.\ref{fig1}(a)).
The conventional Bloch QGT, however, is an $E1$--$E1$ object constructed
from two electric-dipole matrix elements (Fig.\ref{fig1}(b)). This raises a broader question: can each electromagnetic
response channel be assigned its own QGT, pairing its
gauge-field-space response curvature with a quantum metric?
Such a construction would place the Bloch QGT within a broader
electromagnetic hierarchy and connect its members through
nonlocal response.

The conventional Bloch QGT is local in momentum and describes vertical
transitions at a fixed crystal momentum. A field carrying momentum
$\mathbf q$ instead connects states at $\mathbf{k}-\mathbf{q}/2$ and
$\mathbf{k}+\mathbf{q}/2$. Such nonvertical transitions govern optical
spatial dispersion and natural optical activity
\cite{MalashevichSouza2010,MineevYoshioka2010,MaPesin2015,
ZhongMooreSouza2016,RouEtAl2017,PozoSouza2023},
finite-wave-vector Hall response
\cite{Harper2018,Kozii2021},
and general inhomogeneous electrodynamics
\cite{McKay2024}.
Related phenomena include the anomalous skin effect, Pippard
superconductivity, and orbital response to nonuniform magnetic fields
\cite{ReuterSondheimer1948,Pippard1953,KoshinoEtAl2009,
ArimuraAndo2012},
photon drag and spatially structured photocurrents
\cite{GlazovGanichev2014,ShiEtAl2021,Gunyaga2023,
XieNagaosa2025,Qu2025,RofiiHasdeo2026,
GassnerChaudharyClaassen2026}, electron-phonon interaction\cite{HWang2025},
and finite-wavelength superconducting response
\cite{PorllesChen2023,Wei2025}.
Existing formulations capture complementary limits: the uniform orbital
magnetoelectric response describes $\mathbf q=0$
\cite{MalashevichEtAl2010,MalashevichSouza2010},
multipole theories organize the first powers of $\mathbf q$
\cite{PozoSouza2023,DuffSipe2025},
and a recent many-body construction identifies the finite-$\mathbf q$
current--current metric and its optical sum rule
\cite{OnishiFu2026}.
Most spatial-dispersion theories nevertheless take a response-first route:
they expand a finite-wave-vector response and then identify conventional
quantum-geometric quantities in its coefficients
\cite{Kozii2021,PozoSouza2023}.
A common gauge-field space formulation is therefore needed to
associate these response channels with their full QGTs,
extend their geometry to finite momentum, and connect the
local and nonlocal hierarchies.

Here we construct a hierarchy of electromagnetic QGTs in
gauge-field space. For each channel, the intrinsic antisymmetric
adiabatic response is a Berry curvature, and the specified
field-dependent projector defines a unique gauge-field QGT.
This principle applies to both uniform and spatially varying
fields, yielding hierarchies of local and nonlocal quantum
geometries. As an explicit example, we construct the
finite-$\mathbf{q}$ current--current QGT connecting Bloch
states at $\mathbf{k}\pm\mathbf{q}/2$ (Fig.~\ref{fig1}(c)).
Viewed as a function of \(\mathbf q\), the finite-momentum current–current QGT serves as a generating tensor: its normal Taylor coefficients contain combinations of uniform electromagnetic multipole-pair QGTs, whereas tangential derivatives give momentum-space QGT multipoles (Fig.~\ref{fig1}(d))~\cite{Law2023}. We derive the first two normal orders explicitly and connect the resulting curvature and metric to intrinsic nonlocal Hall response and symmetric optical absorption, respectively.

\noindent \textit{\textcolor{blue}{Hierarchy of electromagnetic quantum geometries.}}---
We set $e=\hbar=1$ and work with quantities per unit volume, so that $V=1$. Electromagnetic couplings naturally define a parameter space whose coordinates are field amplitudes conjugate to the current, magnetic moment, and higher electric and magnetic multipole observables. A physical driving protocol specifies a trajectory in this gauge-field space, whereas the quantum geometry characterizes the field-dependent state itself. Denoting these coordinates collectively by $\lambda_I$, we consider a smooth Hermitian family $H(\lambda)$ and let $P(\lambda)$ project onto a smooth isolated state or spectral subspace.
Define the projected deformation operator
$\mathcal{T}_I=(1-P)(\partial_I P)P$, where
$\partial_I\equiv\partial/\partial\lambda_I$.
It maps the selected subspace into its orthogonal complement
and describes its deformation under field $\lambda_I$.
The gauge-field QGT is~\cite{ProvostVallee1980,Berry1984,Kolodrubetz2017}
\begin{equation}
Q_{IJ}[P]
=\operatorname{Tr}\!\left[\mathcal{T}_I^\dagger\mathcal{T}_J\right]
=\operatorname{Tr}\!\left[P(\partial_I P)(1-P)(\partial_J P)\right].
\label{eq1}
\end{equation}
For a nondegenerate state $|n(\lambda)\rangle$, let
$P_n=|n\rangle\langle n|$ and
$\mathcal{T}_{I,n}=(1-P_n)(\partial_I P_n)P_n
=(1-P_n)|\partial_I n\rangle\langle n|$.
Then
\begin{equation}
Q_{IJ}^{(n)}=\operatorname{Tr}\!\left[
\mathcal{T}_{I,n}^\dagger\mathcal{T}_{J,n}\right]
=\langle\partial_I n|(1-P_n)|\partial_J n\rangle.
\label{eq2}
\end{equation}

The generalized force conjugate to $\lambda_I$ is
\begin{equation}
\hat O_I=-\partial_I H.
\label{eq3}
\end{equation}
All field derivatives refer to the same Hamiltonian family and boundary conditions.

The real and imaginary parts of the QGT in Eq.~(\ref{eq1}) are~\cite{ProvostVallee1980,Berry1984,XiaoChangNiu2010,Kolodrubetz2017},
\begin{equation}
g_{IJ}
=
\operatorname{Re}Q_{IJ},
~~
\mathcal{F}_{IJ}
=
-2\operatorname{Im}Q_{IJ}
=
i\operatorname{Tr}
\left\{
P[\partial_I P,\partial_J P]
\right\}.\label{eq4}
\end{equation}
The curvature and fidelity interpretations are derived in
Sec.~I of the Supplemental Material (SM)~\cite{SM}.
For slowly varying fields, adiabatic perturbation theory gives~\cite{Berry1984,XiaoChangNiu2010,Kolodrubetz2017}
\begin{equation}
\langle\hat O_I\rangle
=
-\partial_I E_0
+
\mathcal{F}_{IJ}\dot{\lambda}_J
+
O(\dot{\boldsymbol{\lambda}}^{\,2},
\ddot{\boldsymbol{\lambda}}).\label{eq5}
\end{equation}
The derivation is given in SM, Sec.~V~A.
Equation (\ref{eq5}) identifies ${\cal F}_{IJ}$ with the intrinsic
antisymmetric adiabatic response of $\hat O_I$ to
$\dot\lambda_J$.

The gauge-field QGT is a response geometry: its coordinates are
external gauge-field amplitudes rather than the crystal momentum
$\mathbf k$ of the conventional Bloch QGT, and its curvature gives
the intrinsic adiabatic response to the corresponding field rates.
Its construction therefore requires $H_0$ together with the
perturbation $H'$, which specifies how the fields couple to the
system; the resulting field-dependent projector uniquely fixes the
tensor. This construction applies to the finite-momentum QGT $Q^{AA}(\bf q)$
and to all uniform multipole-pair QGTs. At zero momentum, the uniform current--current
block $Q^{AA}(0)$ is exceptional because its current vertices make
the field-space tensor coincide with the conventional
$\mathbf k$-space Bloch QGT as discussed in Eq.~(\ref{eq9}).

Choosing electromagnetic-multipole amplitudes
$\lambda^{X\ell}_{\alpha}$ as field coordinates generates
the blocks
$Q^{X\ell,Y\ell'}_{\alpha\beta}
=\operatorname{Tr}[
(\mathcal T^{X\ell}_{\alpha})^\dagger
\mathcal T^{Y\ell'}_{\beta}]$,
where $X,Y\in\{E,M\}$, $\ell,\ell'$ denote multipole orders,
and $\alpha,\beta$ collect Cartesian indices.
These blocks form a hierarchy of gauge-field QGTs for electric, magnetic, and higher-multipole response channels, whose curvature components give the field-space curl of the Berry connection and whose metric components determine the quantum-distance line element between nearby states [Eq. (\ref{eq4})].

At the lowest multipole orders, we identify the $E1$ field coordinate
with the uniform vector-potential amplitude,
$\lambda_a^{E1}\equiv A_a$, and the $M1$ field coordinate with a
magnetic-channel amplitude,
$\lambda_\rho^{M1}\equiv\eta_\rho^{B}$. The purpose of introducing $\eta^B$ is to make sure that $A_a$ and $\eta^B_\rho$ are independent variables. We use the shorthand
$\partial_a^{A}\equiv\partial/\partial A_a$ and
$\partial_\rho^{B}\equiv\partial/\partial\eta_\rho^{B}$. To linear
order in the fields,
\begin{equation}
\begin{aligned}
H(A,\eta^{B})
&=
H_0-A_a\hat J_a-\eta_\rho^{B}\hat\mu_\rho+O(\lambda^2),\\
\hat J_a
&=
-\left.\partial_a^{A}H\right|_{0},
\qquad
\hat\mu_\rho
=
-\left.\partial_\rho^{B}H\right|_{0},
\end{aligned}
\label{eq6}
\end{equation}
where the subscript $0$ denotes $A=\eta^B=0$, and $\hat\mu_\rho$ is the orbital magnetic moment operator.

The corresponding gauge-field space QGT has the block structure
\begin{equation}
Q^{\mathrm{EM}}
=
\begin{pmatrix}
Q^{AA}_{ab} & Q^{AB}_{a\rho}\\
Q^{BA}_{\sigma b} & Q^{BB}_{\sigma\rho}
\end{pmatrix},
\qquad
Q^{BA}_{\rho a}
=
\left(Q^{AB}_{a\rho}\right)^{*}.
\label{eq7}
\end{equation}
Here the $AA$, $AB$, $BA$, and $BB$ blocks describe current--current, current--magnetization, magnetization--current, and magnetization--magnetization geometry, respectively. Equation~(\ref{eq5}) gives
\begin{equation}
\begin{pmatrix}
\delta\langle\hat J_a\rangle\\
\delta\langle\hat\mu_\sigma\rangle
\end{pmatrix}_{\mathrm{geom}}
=
\begin{pmatrix}
{\cal F}^{AA}_{ab} & {\cal F}^{AB}_{a\rho}\\
{\cal F}^{BA}_{\sigma b} & {\cal F}^{BB}_{\sigma\rho}
\end{pmatrix}
\begin{pmatrix}
\dot A_b\\
\dot\eta_\rho^{B}
\end{pmatrix}.
\label{eq8}
\end{equation}
Here the first superscript index $X$ in ${\cal F}^{XY}$ labels the measured
generalized force and the second driving field. For a uniform
electric-field drive, $\dot A_b=-E_b$ and $\dot\eta_\rho^B=0$, so
Eq.~(\ref{eq8}) gives
$\delta\langle\hat J_a\rangle_{\rm geom}=-{\cal F}^{AA}_{ab}E_b$ and
$\delta\langle\hat\mu_\rho\rangle_{\rm geom}
=-{\cal F}^{BA}_{\rho b}E_b$.
Thus, ${\cal F}^{AA}$ and ${\cal F}^{BA}$ govern electric-field-induced
current and orbital magnetization, respectively, whereas ${\cal F}^{AB}$
and ${\cal F}^{BB}$ analogously govern responses to a time-dependent
magnetic field. The bulk ${\cal F}^{BA}$ response can also be evaluated
using the semiclassical wave-packet theory of Xiao \textit{et al.}
~\cite{Xiao2021}.

For a noninteracting Bloch Hamiltonian, the uniform $AA$ block is
\begin{eqnarray}
Q^{AA}_{ab,n}(\mathbf{k})
&=&\sum_{m\ne n} {\cal T}^A_{a,nm} {\cal T}^A_{b,mn} =
\sum_{m\ne n}
\frac{
J^a_{nm}(\mathbf{k})J^b_{mn}(\mathbf{k})
}{
\Delta_{mn}^{2}(\mathbf{k})
}\nonumber\\
&=&
\sum_{m\ne n}
r^a_{nm}(\mathbf{k})r^b_{mn}(\mathbf{k})
\equiv
Q^{\mathrm{Bloch}}_{ab,n}(\mathbf{k}),
\label{eq9}
\end{eqnarray}
where
$J^a_{mn}(\mathbf{k})
=i\Delta_{mn}(\mathbf{k})r^a_{mn}(\mathbf{k})$ and $\Delta_{mn} = \epsilon_m - \epsilon_n$.

\noindent \textit{\textcolor{blue}{Finite-momentum gauge-field QGT.}}---
To extend this construction to nonlocal response, consider a
monochromatic vector potential
$A_a(\mathbf q,t)=A_a(\mathbf q,\omega)e^{-i\omega t}$.
At fixed time, the field amplitudes define the Hamiltonian family
\begin{equation}
H[A](t)=H_0-A_a(\mathbf q,t)\hat J_a(-\mathbf q)
              -A_a(-\mathbf q,t)\hat J_a(\mathbf q)+O(A^2).
\end{equation}
For a real field,
$A_a(-\mathbf q,t)=A_a(\mathbf q,t)^{*}$ and
$\hat J_a(-\mathbf q)^\dagger=\hat J_a(\mathbf q)$.
The QGT is evaluated at $\mathbf A=0$, using
$\operatorname{Re}A_a(\mathbf q)$ and
$\operatorname{Im}A_a(\mathbf q)$ as independent field coordinates. The frequency enters only through the time dependence;
in temporal gauge,
$E_a(\mathbf q,\omega)=i\omega A_a(\mathbf q,\omega)$.
The distinction between static transverse sources and the finite-frequency electric-field limit is discussed in SM, Sec. VI B.

For a noninteracting Bloch Hamiltonian, let
$|\psi_{n\mathbf k}\rangle
=e^{i\mathbf k\cdot\mathbf r}|u_{n\mathbf k}\rangle$
be a Bloch eigenstate with energy $\varepsilon_n(\mathbf k)$, and
define
\begin{equation}
\mathbf k_\pm
=
\mathbf k\pm\frac{\mathbf q}{2},
\qquad
\Delta_{mn}(\mathbf k,\mathbf q)
=
\varepsilon_m(\mathbf k_+)-\varepsilon_n(\mathbf k_-).
\end{equation}
We choose the conserved finite-momentum current~\cite{McKay2024}
\begin{align}
\hat J_a(-\mathbf q)
&=\sum_i\int_0^1 d\lambda\,
e^{i(1-\lambda)\mathbf q\cdot\hat{\mathbf r}_i}
\hat v_{ia}
e^{i\lambda\mathbf q\cdot\hat{\mathbf r}_i}
\notag\\
&=\sum_{mn}\int_{\mathbf k}
j^a_{mn}(\mathbf k,\mathbf q)\,
c^\dagger_{m,\mathbf k_+}c_{n,\mathbf k_-},
\end{align}
where
$\int_{\mathbf k}\equiv
\int_{\mathrm{BZ}}d^d k/(2\pi)^d$ and
\begin{equation}
j^a_{mn}(\mathbf k,\mathbf q)
=\int_{-1/2}^{1/2}ds\,
\left\langle u_{m,\mathbf k_+}\right|
\partial_{k_a}H(\mathbf k+s\mathbf q)
\left|u_{n,\mathbf k_-}\right\rangle ,\label{eq13-1}
\end{equation}
where $\hat v_{ia}=i[\hat H_0,\hat r_{ia}]$, $\mathbf q$ is the momentum transferred by the gauge field,
whereas $\mathbf k$ labels the center momentum of the transition.
Hermiticity implies
\begin{equation}
\left[j^a_{mn}(\mathbf k,\mathbf q)\right]^*
=
j^a_{nm}(\mathbf k,-\mathbf q).
\end{equation}

Resolving the projected deformation into transitions from band $n$
gives (SM, Sec.~II~A)
\begin{equation}
\mathcal T^A_{a,n}(\mathbf k;\mathbf q)
=\sum_{m\ne n}
\frac{j^a_{mn}(\mathbf k,\mathbf q)}
{\Delta_{mn}(\mathbf k,\mathbf q)}
|u_{m,\mathbf k_+}\rangle
\langle u_{n,\mathbf k_-}|.
\end{equation}
The corresponding current--current QGT is
\begin{eqnarray}
Q^{AA}_{ab,n}(\mathbf k;\mathbf q)
&=&\operatorname{Tr}\!\left[
(\mathcal T^A_{a,n})^\dagger\mathcal T^A_{b,n}\right]\nonumber\\
&=&\sum_{m\ne n}
\frac{j^a_{nm}(\mathbf k,-\mathbf q)
      j^b_{mn}(\mathbf k,\mathbf q)}
{\Delta_{mn}^2(\mathbf k,\mathbf q)}.\label{eq10}
\end{eqnarray}

To simplify the notation, we henceforth write $Q_{ab,n}(\mathbf{k};\mathbf{q})
\equiv Q^{AA}_{ab,n}(\mathbf{k};\mathbf{q})$. Although $Q_{ab}(\mathbf{k};\mathbf{q})$ is Hermitian in the
polarization indices at fixed $\mathbf{q}$,
\begin{equation}
Q_{ba}(\mathbf{k};\mathbf{q})
=
Q_{ab}(\mathbf{k};\mathbf{q})^{*},
\end{equation}
its real and imaginary parts are not, by themselves, the physical
metric and curvature at finite momentum. Because the Fourier
amplitudes at $\mathbf{q}$ and $-\mathbf{q}$ are related by reality,
the physical kernels require the geometric transpose. For ground-state fidelity and response, we use the occupied
projector: in Eq.~(\ref{eq10}), $n$ is occupied and the final-state
sum is restricted to unoccupied bands (SM, Sec.~II~D). Writing
\begin{equation}
\overline{Q}_{ab}(\mathbf{q})
=
\sum_{n\in\mathrm{occ}}\int_{\mathbf{k}}
Q_{ab,n}(\mathbf{k};\mathbf{q}),
\end{equation}
we define
\begin{align}
g_{ab}(\mathbf{q})
&=
\frac{1}{2}
\left[
\overline{Q}_{ab}(\mathbf{q})
+
\overline{Q}_{ba}(-\mathbf{q})
\right],
\nonumber\\
{\cal F}_{ab}(\mathbf{q})
&=
i\left[
\overline{Q}_{ab}(\mathbf{q})
-
\overline{Q}_{ba}(-\mathbf{q})
\right].\label{eq11}
\end{align}
Equation~(\ref{eq11}) converts the tensor defined for complex Fourier
amplitudes into the physical metric and curvature in real-field
coordinates. It implies
$g_{ba}(-\mathbf q)=g_{ab}(\mathbf q)$ and
$F_{ba}(-\mathbf q)=-F_{ab}(\mathbf q)$. Each transition
contribution $Q_{ab,n}(\mathbf k;\mathbf q)$ is invariant under
independent Bloch-gauge transformations at $\mathbf k_\pm$.
The real-coordinate construction, fidelity interpretation, and
occupied-projector formulation are detailed in SM, Secs.~II B--D.
The finite-momentum current–current QGT, with the real-field prescription in Eq. (\ref{eq11}), thus generates the long-wavelength response geometry through its derivatives with respect to \(\mathbf q\).

\noindent \textit{\textcolor{blue}{Tangential and normal variations.}}---
The pair $(\mathbf{k}_{+},\mathbf{k}_{-})$, with
$\mathbf{k}_{\pm}=\mathbf{k}\pm\mathbf{q}/2$, defines a point in the
doubled Brillouin zone
$\mathrm{BZ}^{(2)}=\mathrm{BZ}_{+}\times\mathrm{BZ}_{-}$ as illustrated in Fig.~\ref{fig1}(d).
Vertical transitions form the diagonal submanifold
$\mathcal{D}:\mathbf{k}_{+}=\mathbf{k}_{-}$, on which
$Q_{ab,n}(\mathbf{k};0)=Q^{\mathrm{Bloch}}_{ab,n}(\mathbf{k})$.
The natural product metric is
\begin{equation}
d\ell^{2}
=d\mathbf{k}_{+}^{2}+d\mathbf{k}_{-}^{2}
=2d\mathbf{k}^{2}+\frac{1}{2}d\mathbf{q}^{2}.
\label{eq12}
\end{equation}
Thus, $\partial_{k_c}$ is tangential to $\mathcal{D}$ and moves the
two transition endpoints together, whereas $\partial_{q_c}$ is normal
to $\mathcal{D}$ and changes their separation at fixed center
momentum.

For $N\geq0$, the tangential derivatives are
\begin{equation}
Q^{k(N)}_{ab,n|c_{1}\cdots c_{N}}(\mathbf{k})
=
\partial_{k_{c_{1}}}\cdots\partial_{k_{c_{N}}}
Q_{ab,n}(\mathbf{k};\mathbf{0}).
\label{eq13}
\end{equation}
They are determined entirely by the diagonal Bloch QGT. Their metric
and curvature parts give the conventional quantum-metric and
Berry-curvature multipoles.

The normal derivatives are
\begin{equation}
Q^{\perp(N)}_{ab,n|c_{1}\cdots c_{N}}(\mathbf{k})
=
\left.
\partial_{q_{c_{1}}}\cdots\partial_{q_{c_{N}}}
Q_{ab,n}(\mathbf{k};\mathbf{q})
\right|_{\mathbf{q}=0}.
\label{eq14}
\end{equation}
For $N\geq1$, these coefficients probe nonvertical
deformations that are not determined by the diagonal
Bloch QGT. Their multipole decomposition connects
finite-momentum geometry to the hierarchy of uniform
gauge-field space QGTs.

Differentiating Eq.(\ref{eq11}), each derivative acting
on the $-\mathbf q$ argument contributes a minus sign.
The physical metric and curvature coefficients therefore
depend on the parity of the normal order:
\begin{align}
g^{\perp(N)}
&=\operatorname{Re}Q^{\perp(N)},
&
F^{\perp(N)}
&=-2\operatorname{Im}Q^{\perp(N)},
&& N\ {\rm even},
\nonumber\\
g^{\perp(N)}
&=i\operatorname{Im}Q^{\perp(N)},
&
F^{\perp(N)}
&=2i\operatorname{Re}Q^{\perp(N)},
&& N\ {\rm odd}.\label{eq16}
\end{align}
Odd-order coefficients are purely imaginary in momentum
space; their real-space gradient coefficients are obtained
by multiplying by $i^{-N}$.

\begin{figure}[t!]
\includegraphics[width=0.9\columnwidth]{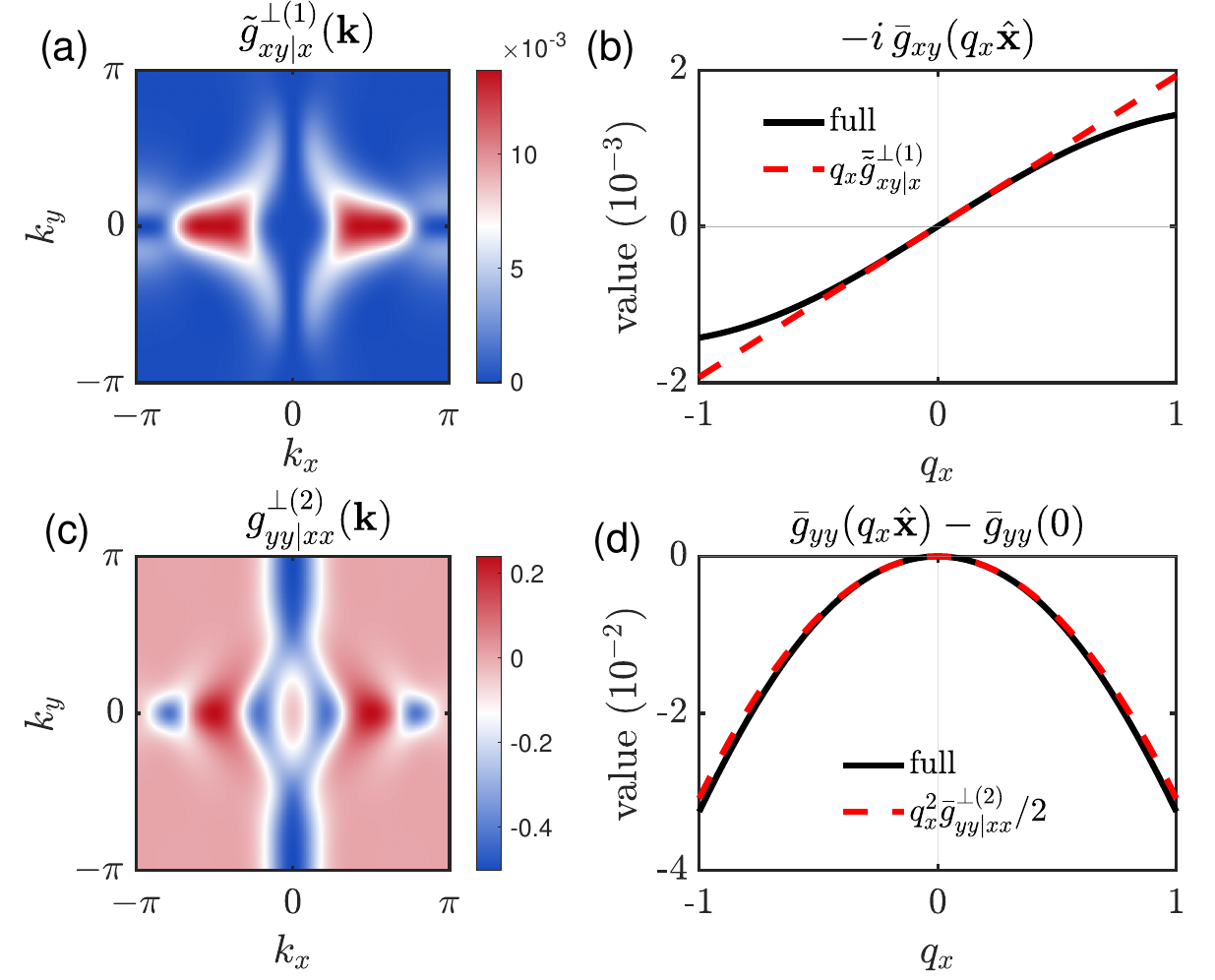}
\caption{Normal expansion for the model in the text. (a,c) Brillouin-zone distributions of $\widetilde g^{\perp(1)}_{xy|x}(\mathbf k)\equiv-i\sum_{n\in\mathrm{occ}}g^{\perp(1)}_{xy|x,n}(\mathbf k)$ and $g^{\perp(2)}_{yy|xx}(\mathbf k)\equiv\sum_{n\in\mathrm{occ}}g^{\perp(2)}_{yy|xx,n}(\mathbf k)$. (b,d) Exact Brillouin-zone-integrated metrics (solid) and their leading normal expansions (dashed).}
\label{fig2}
\end{figure}

To resolve their electromagnetic content, expand the
projected deformation operator as
\begin{equation}
\mathcal T_a^A(\mathbf q)
=\mathcal T_a^{E1}
+iq_c\mathcal S_{ac}
-q_cq_d\mathcal U_{acd}+O(q^3),\label{eq24-1}
\end{equation}
where
\begin{equation}
\begin{aligned}
\mathcal S_{ac}
&=\mathcal T^{E2}_{ac}
+\epsilon_{\rho ca}\mathcal T^{M1}_{\rho},\\
\mathcal U_{acd}
&=\mathcal T^{E3}_{acd}
+{\rm Sym}_{cd}\epsilon_{\rho ca}
 \mathcal T^{M2}_{\rho d},
\end{aligned}
\end{equation}
and ${\rm Sym}_{cd}X_{cd}=(X_{cd}+X_{dc})/2$. Equation (\ref{eq24-1}) is illustrated schematically in Fig.\ref{fig1}(a).
Here $\mathbf k$ and $n$ are suppressed, and each
multipole deformation operator is defined by its
corresponding field derivative. The field profiles
and normalization conventions are given in SM Sec.III A-C.
The multipole-pair QGTs are overlaps of the corresponding
deformation operators; for example,
\begin{equation}
Q^{E1,E2}_{a,bc}
=\operatorname{Tr}\!\left[
(\mathcal T^{E1}_a)^\dagger\mathcal T^{E2}_{bc}\right],
\qquad
Q^{Y,X}_{\beta,\alpha}
=(Q^{X,Y}_{\alpha,\beta})^*.
\end{equation}
Expanding $Q^{AA}_{ab}
=\operatorname{Tr}[(\mathcal T_a^A)^\dagger\mathcal T_b^A]$
gives
\begin{equation}
\begin{aligned}
Q^{\perp(1)}_{ab|c}
=i\Big[
& Q^{E1,E2}_{a,bc}
-\left(Q^{E1,E2}_{b,ac}\right)^{*} \\
&+\epsilon_{\rho cb}
 \left(Q^{M1,E1}_{\rho,a}\right)^{*}
-\epsilon_{\rho ca}Q^{M1,E1}_{\rho,b}
\Big].
\end{aligned}
\label{eq17}
\end{equation}
The exact second-order identity is
\begin{equation}
\begin{aligned}
Q^{\perp(2)}_{ab|cd}
&=-2\left[
Q^{E1,E3}_{a,bcd}
+\left(Q^{E1,E3}_{b,acd}\right)^{*}
\right] \\
&+2 {\rm{Sym}}_{cd}\Big[
Q^{E2,E2}_{ac,bd}
+\epsilon_{\rho ca}Q^{M1,E2}_{\rho,bd}
+\epsilon_{\rho db}
 \left(Q^{M1,E2}_{\rho,ac}\right)^{*} \\
&
+\epsilon_{\rho ca}\epsilon_{\sigma db}
 Q^{M1,M1}_{\rho,\sigma}
-\epsilon_{\rho cb}Q^{E1,M2}_{a,\rho d}
-\epsilon_{\rho ca}
 \left(Q^{E1,M2}_{b,\rho d}\right)^{*}
\Big].
\end{aligned}
\label{eq18}
\end{equation}
Equations (\ref{eq17}) and (\ref{eq18}) make the generating relation explicit. The first normal coefficient combines electric-dipole–quadrupole and magnetic-dipole–electric-dipole geometries; the second additionally involves electric octupole, electric-quadrupole-pair, magnetic-dipole-pair, and magnetic-quadrupole geometries.

\begin{table}[t]
\centering
\caption{Physical response channels associated with the multipole-pair curvatures in Eq. (\ref{eq19}). The second column identifies the measured generalized force, and the third gives the driving field rate.
The symmetric amplitude
$\lambda^{E2}_{bd}=\lambda^{E2}_{db}$ couples to the electric quadrupole; the $M1$ channel refers to the orbital magnetic
moment, while $M2$ denotes the full traceless magnetic-gradient
channel (SM, Secs. III B--D). Field normalizations follow
the SM conventions. The cited works study the corresponding response channels, although they were not previously identified as Berry curvatures.}
\label{tab:multipole_curvatures}
\begin{tabular}{llc}
\hline\hline
Coefficient & Measured channel & Driving field rate \\
\hline
$\mathcal Q_{bc,a}$~\cite{MalashevichSouza2010,MahonSipe2020}
& Electric quadrupole ($E2$)
& $\dot A_a$ \\

$\mathcal M_{\rho,a}$~\cite{Xiao2021,MahonSipe2020}
& Magnetic dipole ($M1$)
& $\dot A_a$ \\

$\mathcal O_{bcd,a}$~\cite{DuffSipe2025}
& Electric octupole ($E3$)
& $\dot A_a$ \\

$\mathcal G_{ac,bd}$~\cite{DuffSipe2025}
& Electric quadrupole ($E2$)
& $\dot\lambda^{E2}_{bd}$ \\

$\mathcal C_{\rho,bd}$~\cite{DuffSipe2025}
& Magnetic dipole ($M1$)
& $\dot\lambda^{E2}_{bd}$ \\

$\mathcal B_{\rho,\sigma}$~\cite{DuffSipe2025}
& Magnetic dipole ($M1$)
& $\dot\eta^B_\sigma$ \\

$\mathcal N_{\rho d,a}$~\cite{DuffSipe2025}
& Magnetic quadrupole ($M2$)
& $\dot A_a$ \\
\hline\hline
\end{tabular}
\end{table}

By Eq. (\ref{eq5}), the multipole-pair curvatures are intrinsic adiabatic-response coefficients. We denote them by
\begin{eqnarray}
\mathcal{Q}_{bc,a}
&\equiv& {\cal F}^{E2,E1}_{bc,a},~~
\mathcal{M}_{\rho,a}
\equiv {\cal F}^{M1,E1}_{\rho,a},~~
\mathcal{O}_{bcd,a}
\equiv {\cal F}^{E3,E1}_{bcd,a},\nonumber\\
\mathcal{G}_{ac,bd}
&\equiv& {\cal F}^{E2,E2}_{ac,bd},~~
\mathcal{C}_{\rho,bd}
\equiv {\cal F}^{M1,E2}_{\rho,bd},~~
\mathcal{B}_{\rho,\sigma}
\equiv {\cal F}^{M1,M1}_{\rho,\sigma},\nonumber
\\
\mathcal{N}_{\rho d,a}
&\equiv& {\cal F}^{M2,E1}_{\rho d,a}.
\label{eq19}
\end{eqnarray}
The measured channels and driving field rates associated
with these curvature coefficients are summarized in TABLE I.
The full QGT blocks additionally contain
the complementary quantum metrics.


\textit{Finite-momentum Hall response.---} Using
$\dot A_b(\mathbf q)=-E_b(\mathbf q)$ in Eq.~(\ref{eq5}), the intrinsic
Hall conductivity is
\[
\sigma^H_{ab}(\mathbf q)
=-\sum_{n\in{\rm occ}}\int_{\mathbf k}
F_{ab,n}(\mathbf k;\mathbf q).
\]
Its normal expansion resolves the nonlocal response into
multipole-pair curvatures (SM, Sec.IV and Sec.~VI~A--B). For the
time-reversal-symmetric model below, the integrated adiabatic
curvature vanishes; the numerical illustration therefore
focuses on the metric hierarchy.

\textit{Symmetric optical absorption.---}
The metric sector is accessible through an optical
sum rule. Let $\sigma^{\rm abs}_{ab}(\mathbf q,\omega)$
denote the absorptive conductivity and define its
index-symmetric part as
$\sigma^S_{ab}=(\sigma^{\rm abs}_{ab}
+\sigma^{\rm abs}_{ba})/2$.
For the occupied-projector QGT in the time-reversal-symmetric
setting considered here, the sum rule reads
\begin{eqnarray}
\frac{1}{\pi}\int_0^\infty\frac{d\omega}{\omega}\,
\operatorname{Re}\sigma^{S}_{ab}(\mathbf q,\omega)
&=&
\frac{g_{ab}(\mathbf q)+g_{ba}(\mathbf q)}{2}\label{eq20}\\
&=&
g_{ab}(\mathbf 0)
+\frac{1}{2}q_cq_d\,g^{\perp(2)}_{ab|cd}
+O(q^4).\nonumber
\end{eqnarray}
The derivation and the general two-direction sum rule are
given in SM, Secs.~VI~D--E.

\noindent \textit{\textcolor{blue}{Numerical illustration.}}---
Figure~\ref{fig2} illustrates the
first- and second-order normal expansions of the finite-momentum
quantum metric in a time-reversal-symmetric, inversion-broken
insulator. The leading odd-in-$\mathbf q$ off-diagonal and
even-in-$\mathbf q$ diagonal terms reproduce the corresponding
full kernels at small $\mathbf q$. The model, its symmetry constraints, and the normal-expansion results are detailed in the End Matter; current vertices, numerical implementation, and convergence tests are given in SM, Sec. VII. Finally, possible experimental detection of $Q^{AA}({\bf q})$ is discussed in End Matter.

\noindent \textit{\textcolor{blue}{Conclusion.}}---
The field-dependent projector defines a hierarchy of electromagnetic multipole-pair QGTs, with the conventional Bloch QGT as its uniform E1–E1 member. The normal expansion of the finite-momentum current–current QGT organizes combinations of these geometries, connecting their curvature and metric sectors to nonlocal Hall response and an inverse-frequency moment of symmetric optical absorption, respectively.

\bigskip
We acknowledge support from the National Natural Science Foundation of China (Grants No. 12674063, No. 12574054, and No. 12474047).

\newpage
\onecolumngrid
\section*{End Matter}
\par

\twocolumngrid

\bigskip

\noindent\textit{\textcolor{blue}{Numerical illustration of the normal expansion. }---}
We illustrate the finite-momentum geometry using the
time-reversal-symmetric, inversion-broken Hamiltonian
\begin{equation}
\begin{aligned}
H(\mathbf k)
={}&a\sin k_x\,\tau_xs_z+a\sin k_y\,\tau_y\\
&+\left[m-b(2-\cos k_x-\cos k_y)\right]\tau_z\\
&+\lambda_R(\sin k_y\,s_x-\sin k_x\,s_y)
+\delta\tau_x ,
\end{aligned}
\label{eq:EM_numerical_model}
\end{equation}
where $\tau_i$ and $s_i$ are Pauli matrices acting in
orbital and spin space, respectively; identity matrices
are implicit. The lattice constant is set to unity,
and the Brillouin zone is $[-\pi,\pi)^2$.
We take $a=b=m=1$, $\lambda_R=0.5$, and $\delta=0.2$,
with the chemical potential in the insulating gap.
The parameter $\delta$ breaks inversion symmetry while
preserving time-reversal symmetry.

For momentum transfer $\mathbf q=q_x\hat{\mathbf x}$, the
transition endpoints are $\mathbf k_\pm=\mathbf k\pm\mathbf q/2$.
For each $\mathbf k$, we evaluate $Q_{ab,n}(\mathbf k;\mathbf q)$
from Eq.~(\ref{eq10}) and construct the metric kernel
$g_{ab,n}(\mathbf k;\mathbf q)$ using the geometric-transpose
combination in Eq.~(\ref{eq11}). Summing over occupied bands and
integrating over the Brillouin zone gives
\begin{equation}
\bar g_{ab}(\mathbf q)
=\sum_{n\in{\rm occ}}\int_{\rm BZ}
\frac{d^2k}{(2\pi)^2}\,
g_{ab,n}(\mathbf k;\mathbf q).
\label{eq:EM_integrated_metric}
\end{equation}

The Hamiltonian in Eq.~(\ref{eq:EM_numerical_model}) also preserves the mirror
$\mathcal M_x=i s_x$, satisfying
$\mathcal M_xH(k_x,k_y)\mathcal M_x^{-1}=H(-k_x,k_y)$.
Under this mirror, $q_x\to-q_x$, $A_x\to-A_x$, and
$A_y\to A_y$. Consequently,
\[
\bar g_{xy}(q_x\hat{\mathbf x})
=-\bar g_{xy}(-q_x\hat{\mathbf x}),\qquad
\bar g_{yy}(q_x\hat{\mathbf x})
=\bar g_{yy}(-q_x\hat{\mathbf x}).
\]
Their leading normal coefficients are
\begin{equation}
\begin{aligned}
g^{\perp(1)}_{xy|x,n}(\mathbf k)
&=\left.
\partial_{q_x}g_{xy,n}(\mathbf k;\mathbf q)
\right|_{\mathbf q=0},\\
g^{\perp(2)}_{yy|xx,n}(\mathbf k)
&=\left.
\partial_{q_x}^2g_{yy,n}(\mathbf k;\mathbf q)
\right|_{\mathbf q=0}.
\end{aligned}
\label{eq:EM_normal_metric_coefficients}
\end{equation}
By Eq.~(\ref{eq16}), the first-order coefficient is purely imaginary.
We therefore introduce the real quantity
\begin{equation}
\widetilde g^{(1)}_{xy|x}(\mathbf k)
=-i\sum_{n\in{\rm occ}}
g^{\perp(1)}_{xy|x,n}(\mathbf k).
\label{eq:EM_real_first_order_metric}
\end{equation}
Figure~\ref{fig2}(a) displays this quantity, while Fig.~\ref{fig2}(c)
shows $\sum_{n\in{\rm occ}}
g^{\perp(2)}_{yy|xx,n}(\mathbf k)$.

Integrating these coefficients gives the
long-wavelength expansions
\begin{equation}
\begin{aligned}
-i\bar g_{xy}(q_x\hat{\mathbf x})
&=q_x\overline{\widetilde g}^{\perp (1)}_{xy|x}
+O(q_x^3),\\
\bar g_{yy}(q_x\hat{\mathbf x})-\bar g_{yy}(0)
&=\frac{q_x^2}{2}\,
\bar g^{\perp(2)}_{yy|xx}+O(q_x^4),
\end{aligned}
\label{eq24}
\end{equation}
where
\begin{equation}
\begin{aligned}
\overline{\widetilde g}^{\perp (1)}_{xy|x}
&=\int_{\rm BZ}\frac{d^2k}{(2\pi)^2}\,
\widetilde g^{\perp (1)}_{xy|x}(\mathbf k),\\
\bar g^{\perp(2)}_{yy|xx}
&=\sum_{n\in{\rm occ}}\int_{\rm BZ}
\frac{d^2k}{(2\pi)^2}\,
g^{\perp(2)}_{yy|xx,n}(\mathbf k).
\end{aligned}\label{eq25}
\end{equation}
Because $g^{\perp(2)}_{yy|xx}$ is defined as the second
derivative rather than as the quadratic Taylor coefficient,
the second line of Eq.~(\ref{eq24}) contains the factorial
$1/2$; no such factor enters the definition in Eq.~(\ref{eq25}). Through the optical sum rule in Eq.~(\ref{eq20}), the second line of
Eq.~(\ref{eq24}) gives the leading nonlocal correction to the
inverse-frequency moment of the y-polarized transverse absorptive
conductivity $\sigma^{\rm abs}_{yy}$.

Figures~\ref{fig2}(b) and \ref{fig2}(d) compare the full finite-momentum
kernels with the linear and quadratic approximations
in Eq.~\eqref{eq24}, respectively.
Their agreement at small $q_x$ illustrates how the
normal coefficients capture the leading nonlocal
metric contributions. Deviations at larger momentum
reflect higher-order terms beyond the displayed
expansion. The Brillouin-zone discretization, evaluation of the normal
derivatives, and convergence tests are detailed in SM, Sec.~VII.

\noindent \textit{\textcolor{blue}{Experimental access.}}---
Both the curvature and metric sectors of $Q^{AA}$ are accessible
through wave-vector-resolved conductivity. For the curvature, a
periodic gate, interdigitated electrode, or phase-controlled
microwave/terahertz near field can generate an ac longitudinal field
$E_x(q_x,\omega)$. Along the mirror line, source-space antisymmetry
gives
${\cal F}^{AA}_{yx}(q_x,0)\equiv iD^{AA}_{xy}(q_x)$, and hence
\begin{equation}
D^{AA,\mathrm{exp}}_{xy}(q_x)
=
\lim_{\omega\to0^+}
i\frac{J_y(q_x,\omega)}{E_x(q_x,\omega)}.
\label{eq16-1}
\end{equation}
The ratio is formed at finite $\omega$ before taking the
subgap adiabatic limit.
The sheet-current component $J_y(q_x,\omega)$ can be reconstructed in three steps. First, scanning or wide-field nitrogen-vacancy magnetometry maps the stray magnetic field above the sample produced by $J_y(q_x,\omega)$; for an ac drive, phase-resolved detection records both its amplitude and phase relative to $E_x$. Second, the measured magnetic-field map is Fourier transformed. Third, the Fourier-space Biot--Savart relation is inverted to obtain $J_y(q_x,\omega)$~\cite{Chang2017,Tetienne2017,Palm2022,Mizuno2020}.
Using Eq.~(\ref{eq16-1}), $J_y(q_x,\omega)=-iD_{xy}^{AA}(q_x)E_x(q_x,\omega)$. Therefore, $D_{xy}^{AA}(q_x)=d_1q_x$ gives $J_y(q_x,\omega)=-id_1q_xE_x(q_x,\omega)$, where $d_1$ is the linear coefficient. Since $iq_xE_x(q_x,\omega)$ is the Fourier transform of $\partial_xE_x(x,\omega)$, the leading term becomes $J_y(x,\omega)=-d_1\partial_xE_x(x,\omega)$. Thus, the factor $-i$ in momentum space encodes a spatial gradient rather than a dissipative temporal lag.
Extracting
the component odd under $q_x\to-q_x$ and magnetic-domain reversal
suppresses ordinary background currents. Local imaging is especially
useful because this periodic Hall current averages to zero over one
wavelength.

The metric is obtained independently from finite-$q$ absorption.
Measuring
$\sigma^{\mathrm{abs}}_{yy}(q_x,\omega)$ under a
$y$-polarized field and subtracting the $q_x=0$ spectrum gives the second term
$\Delta g_{yy}(q_x)$ in
Eqs.~\eqref{eq20} and (\ref{eq24}). Near-field infrared/terahertz
nanoscopy provides the required momentum resolution and has already
resolved nonlocal optical conductivity in two-dimensional systems
\cite{LundebergEtAl2017}. A finite spectral window yields the
corresponding partial metric weight. Thus the curvature requires a
low-frequency measurement below the finite-$q$ particle--hole gap,
whereas the metric requires an interband-frequency sweep.

\end{document}